# Fermi's Theory of Beta Decay:

# A First Attempt at Electroweak Unification

Luca Nanni

Faculty of Physics, University of Ferrara, Ferrara, Italy

## Abstract

The purpose of this study, mainly historical and pedagogical, is to investigate the physical-mathematical similitudes of the spectroscopic and beta decay Fermi theories. Both theories were formulated using quantum perturbative theory that allowed obtaining equations whose algebraic structure and physical interpretation suggests that the two phenomena occur according to the same mechanism. Fermi, therefore, could have guessed well in advance of the times that the two theories could be unified into a single physical-mathematical model that led to different results depending on the considered energy. The electroweak unification found its full realization only in the 1960s within the Standard Model that, however, is mainly based on a mathematical approach. Retracing the reasoning made by Fermi facilitates the understanding of the physical foundations that underlie the unification of the electromagnetic and weak forces.

## 1  Introduction

In 1933, Fermi attended the Seventh Solvay Conference where the critical points of the atomic physics of the period were debated, including the discovery of the neutron, the positron and the neutrino hypothesis [1]. The physics of these particles was of fundamental importance for explaining the mechanism of nuclear radioactivity which, in the 1930s, had attracted the attention of the most prestigious international research groups [2-3]. In this context, Fermi had been one of the most active and was already known as one of the leading experts in the study of atomic nucleus and artificial radioactivity [4]. With the 1933 conference Fermi had the opportunity to know the state of art of nuclear physics, which, in the same year, allowed him to formulate the first beta decay theory [5]. What became of this monumental work is well known to history: the journal *Nature* declined publication asserting that the theory was based on speculative ideas, forcing Fermi to publish it in the Italian journal *Il Nuovo Cimento*, limiting its international its diffusion [6-7].

Fermi formulated the theory of beta decay following the approach used by Dirac in the formulation of the theory of radiation [8], on the formalism of which Fermi himself had become particularly expert, having published in 1932 another scientific masterpiece which was the spectroscopic theory [9]. But unlike in the latter, where Fermi used the formalism of *classical* quantum mechanics, in the theory of decay, he made use of the formalism of the creation and annihilation operators introduced by Jordan and Wigner some years earlier [10]. It is just this methodological parallelism in the formulation of the two theories and the fact that both correctly explain the experimental results, each for its own field of application, which makes us guess that the two phenomena, apparently so different, have the same physical nature [11-13]. In this context Fermi, in our opinion, had taken the first step towards the electroweak unification. Today, we know that beta decay is mediated by vector bosons and that proton and the neutron are composite particles that differ in the *flavour* of one of the three quarks forming them. Therefore, the physical reality is very different from the Fermi point-contact interaction hypothesis, but it is a fact that Fermi's golden rule, if properly applied, works properly both for spectroscopic and radioactive phenomena (limited at low energies) [14], as well as for the calculation of the tunnelling rate [15]. In this scenario, the two theories of Fermi have the same scientific *dignity* as the monumental and ambitious Standard Model. On the basis of these considerations, it is useful, from an historical and pedagogical perspective, to retrace the fundamental points concerning the formulation of the Fermi beta decay theory, highlighting its analogies with the spectroscopic theory and the similarity of the formalism with the modern electroweak theory.

To simplify the study we take into account negative beta decay:

$$n \to p + e + \bar{\nu}_e, \tag{1}$$

which is compared with the emission of a photon by an atomic electron that decays from an excited state to a lower energetic level:

$$(e)^*_{atomic} \to e_{atomic} + \gamma. \tag{2}$$

The first analogy between the two phenomena appears in the kinematic of their decays, and we begin the study starting from this evidence.

## 2 Kinematics of Radiative and Beta Decays

Around 1930, (negative) beta decay was interpreted as a two-body process similar to alpha decay: a monokinetic reaction leading to a recoiled nucleus and an emitted electron that carries away the entire kinetic energy [16]. However, beta decay provides a wide spectrum of energies which is not compatible with a two-body mechanism, unless to violate the principle of energy conservation. Going back to the scientific orthodoxy of the period, it was necessary to hypothesize a multi-body process that also included unknown particles which the technologies of the time were not yet able to detect. This is what Pauli did with his famous letter of 1930 in which he hypothesized the presence of a neutral particle with a mass lower than that of the electron and with a half-integer spin. This also saves the exchange theorem, not only the principle of energy conservation. This *elusive* particle was called the neutrino by Fermi just during the formulation of his theory on beta decay [17].

Since the kinetic energies of neutron and proton are negligible compare to those of electron and neutrino, and since the neutrino is very light particle (with negligible rest mass energy than that of electron), the squared energy balance of the decay Eq. (1) is:

$$m_n^2 c^4 + p_n^2 c^2 = \left(m_p^2 c^4 + p_p^2 c^2\right) + \left(m_e^2 c^4 + p_e^2 c^2\right) + \left(m_\nu^2 c^4 + p_\nu^2 c^2\right) \stackrel{\cong}{\Rightarrow}$$
$$\stackrel{\cong}{\Rightarrow} m_n^2 c^4 = m_p^2 c^4 + \left(m_e^2 c^4 + p_e^2 c^2\right) + p_\nu^2 c^2 \quad . \quad (3)$$

If the difference between the neutron-proton squared energy mass is $W^2$, Eq. (3) becomes:

$$W^2 = \left(m_e^2 c^4 + p_e^2 c^2\right) + p_\nu^2 c^2, \quad (4)$$

from which we get:

$$\left(p_e^2 + p_\nu^2\right) = \frac{W^2 - m_e^2 c^4}{c^2}. \quad (5)$$

Equation (5) proves that the kinetic energy of the emitted electron is zero when that of the neutrino equals the mass energy variation of the decaying neutron. The impulse of the electron, on the other hand, gets the highest value when the neutrino kinetic energy is zero. Therefore, since the experimental results show that the electron energy is always different from zero, Eq. (5) must comply the following constraint:

$$\sqrt{m_e^2 c^4 + p_e^2 c^2} \leq W. \quad (6)$$

If the neutrino had zero mass, then, at the highest value of the electron impulse, the two members of Eq. (6) would become equal. The experimental results, however, show that this condition is never verified and, therefore, either the neutrino is a massless particle with impulse $h\nu/c$ (with which it can never be null) or must have a non-zero mass. In his original article, Fermi devoted an entire paragraph to the problem of neutrino mass [5] and, therefore, it is reasonable to suppose that he had already taken into consideration the possible results obtained from the study of the dynamics of the process Eq. (1), rejecting the hypothesis of the neutrino as half-integer spin luxon.

Let us consider the radiative emission Eq. (2); indicating with $(m_e c, \boldsymbol{p}^*)$, $(m_e c, \boldsymbol{p})$ the four-impulse of the atomic electron in the starting and final states and with $(h\nu/c, 0)$ that of the photon, the following equations hold:

$$\boldsymbol{p}^{*2} = \boldsymbol{p}^2 + (h\nu/c)^2 \tag{7}$$

$$\Delta \boldsymbol{p}^2 = h\nu \left( \frac{h\nu + 2m_e c^2}{c^2} \right) \ldots \Rightarrow \ldots W_e^2 = h^2 \nu^2 + 2m_e c^2 h\nu . \tag{8}$$

Rearranging Eq. (8), we get the photon impulse:

$$p_\gamma = \frac{\sqrt{W_e^2 - 2m_e c^2 h\nu}}{c}, \tag{9}$$

where $W_e$ is the energy difference between the two states of the radiative process. Equation (9) is analogous to Eq. (5), where the photon plays the role of the electron-neutrino pair and the two different electronic states are equivalent to the states of the neutron-proton couple. The only difference between the two equations is that, while in Eq. (5) there appears the square mass energy of the electron, in Eq. (9) this energy is replaced by a mixed term given by the product between the mass energy of the electron and that of the photon, as to remember that the two particles physically interact with each other. In fact, the photon is a real particle of the radiative processes, while the vector boson $W^-$ is a virtual particle of beta decay, and it is reasonable to expect that it does not appear explicitly in the kinematic equations. Since Fermi approached the theory of beta decay by referring to the spectroscopic theory, it is reasonable to suppose that he had preliminarily made these considerations on the kinetics of the two processes.

## 3      Interaction Hamiltonian

In formulating the beta decay theory, Fermi accepted the neutrino hypothesis and gave up describing light particles within the nucleus. The entire theory is based on the following hypotheses:

- The total number of electrons and neutrinos is not necessarily constant. This is the same hypothesis used by Fermi when formulating the spectroscopic theory about adsorbed and emitted photons.
- The nucleus is formed by protons and neutrons (heavy particles) that interact by exchanging forces, as suggested in the same period by Heisenberg [19] and Majorana [20]. These particles are considered as two quantum states of a single particle that differ in the value of an internal degree of freedom that can take only the following values: +1 for the neutron and –1 for the proton.
- The Hamiltonian of the decay process is given by the contribution of the heavy and the light particles and by an interaction term small enough to be handled with the perturbation method. This Hamiltonian is completed by its complex conjugate representing the opposite process, i.e. positive beta decay. Fermi's idea, therefore, is nothing more than an anticipation of the Feynman-Stueckeberg reinterpretation principle of antimatter [21].

The formalism used by Fermi in beta decay is that of the second quantization [10] and electrons and neutrinos are defined as field operators:

$$\begin{cases} \psi = \sum_i \psi_i a_i \ldots;\ldots \psi^\dagger = \sum_i \psi_i^* a_i^\dagger \ldots \text{for electrons} \\ \varphi = \sum_j \varphi_j b_j \ldots;\ldots \varphi^\dagger = \sum_j \varphi_j^* b_j^\dagger \ldots \text{for neutrinos.} \end{cases} \quad (10)$$

The creator and annihilator operators $a_i^\dagger$, $a_i$, $b_j$ and $b_j^\dagger$ act on the occupation numbers of the electronic states as follows:

$$\begin{cases} a_i \psi(N_1, \cdots N_i, \cdots) = (-1)^{N_1 + \cdots + N_{i-1}} (1 - N_i) \psi(N_1, \cdots N_i, \cdots) \\ a_i^\dagger \psi(N_1, \cdots N_i, \cdots) = (-1)^{N_1 + \cdots + N_{i-1}} N_i \psi(N_1, \cdots N_i, \cdots) \end{cases} \quad (11)$$

$$\begin{cases} b_j \varphi(M_1, \cdots M_j, \cdots) = (-1)^{M_1 + \cdots + M_{j-1}} (1 - M_j) \varphi(M_1, \cdots M_j, \cdots) \\ b_j^\dagger \varphi(M_1, \cdots M_j, \cdots) = (-1)^{M_1 + \cdots + M_{j-1}} M_j \varphi(M_1, \cdots M_j, \cdots). \end{cases} \quad (12)$$

Since electrons and neutrino are fermions, the Pauli Exclusion Principle bounds the possible values of the occupation numbers $N_i$ and $M_j$ to 1 and 0, respectively.

To describe the nuclear heavy particles, Fermi used the Lagrangian formalism (configuration space); each particle is equipped with a set of generalized coordinates and a new internal coordinate $\rho$ upon which acts the operator describing the *neutron → proton* transition. This operator follows:

$$Q = \sigma_x - i\sigma_y \ldots;\ldots Q^\dagger = \sigma_x + i\sigma_y, \quad (13)$$

where $\sigma_i$ are the Pauli matrices; $Q^\dagger$ defines the *neutron → proton* transition, while $Q$ defines the opposite transition. The internal coordinate $\rho$ deals with the quark flavour (*up* versus *down*).

Fermi considered beta decay as a perturbation of the energy system within the nucleus and described its perturbation Hamiltonian equation as:

$$H_{int.}^{(\beta)} = Q \sum_{i,j} c_{ij} a_i b_j + Q^\dagger \sum_{i,j} c_{ij}^\dagger a_i^\dagger b_j^\dagger. \quad (14)$$

In this operator, no electrostatic term due to the interaction between the nucleus and the emitted electron appears. This is because the electrostatic attraction is energetically negligible compared to the energy involved in decay, and because the experimental results prove that it does not contribute to the radioactive phenomenon. The operators $c_{ij}$ and $c_{ij}^\dagger$, whose physical meaning will be explained shortly, depend on the space coordinates, on the impulses and on the internal coordinates.

Using the component spinors of $\psi$ and $\varphi$, Fermi constructed a four-vector $A$ given by the following bilinear combinations:

$$\begin{cases} A_0 = -\psi_1\varphi_2 + \psi_2\varphi_1 + \psi_3\varphi_4 - \psi_4\varphi_3 \\ A_1 = \psi_1\varphi_3 - \psi_2\varphi_4 - \psi_3\varphi_1 + \psi_4\varphi_2 \\ A_2 = i\psi_1\varphi_3 + i\psi_2\varphi_4 - i\psi_3\varphi_1 - i\psi_4\varphi_2 \\ A_3 = -\psi_1\varphi_4 - \psi_2\varphi_3 + \psi_3\varphi_2 + \psi_4\varphi_1 \end{cases} \qquad (15)$$

The four components transform like those of a four-vector, and Fermi interpreted them in analogy to those of the electromagnetic four-potential. The perturbation Hamiltonian Eq. (14) may be therefore rewritten as:

$$H_{int.}^{(\beta)} = g\left(QA_i + Q^*A_i^*\right). \qquad (16)$$

In the approximation that the nucleons are much slower than the emitted light particles, we can consider only the time-component $A_0$:

$$H_{int.}^{(\beta)} = g\left(Q\psi^\dagger\delta\varphi + Q^*\psi\delta\varphi^\dagger\right), \qquad (17)$$

where $\delta$ is the antisymmetric matrix:

$$\delta = \begin{pmatrix} 0 & -1 & & 0 \\ 1 & 0 & & \\ & & 0 & 1 \\ 0 & & -1 & 0 \end{pmatrix}. \qquad (18)$$

The $\delta$ is analogous to that transforming a Dirac spin-up spinor in a spin-down spinor. Comparing Hamiltonian Eq. (16) and Eq. (17), it is clear that:

$$A_0 \Rightarrow \psi^\dagger\delta\varphi. \qquad (19)$$

Equation (19) recalls the correspondence between the quantum electrodynamic components of the four-potential operator $A_\mu$ and the hadronic currents of the modern beta decay theory:

$$A_\mu \Rightarrow \bar{\psi}\gamma_\mu\varphi, \qquad (20)$$

where $\gamma_\mu$ are the Dirac gamma matrices. The quantity g in the Hamiltonian Eq. (16) is a numerical constant that must be determined experimentally, and it depends on the interaction force.

Comparing the Hamiltonian Eqs. (14) and (17), we get the explicit form of the operators $c_{ij}$ and $c_{ij}^\dagger$:

$$c_{ij} = g\psi_i^\dagger\delta\varphi_j \ . \ . \ ; \ . \ . c_{ij}^\dagger = g\psi_i\delta\varphi_j^\dagger. \qquad (21)$$

Since these two operators describe transformations occurring in the nucleus, it follows that the electron $\psi_i$ and neutrino $\varphi_j$ field operators have four components which depend on the nuclear coordinates. Therefore, the values assumed by these operators are significantly different from zero only in the region occupied by the nucleus. This explains why, in the Fermi theory of beta decay, the interaction is a short range force. The two operators in Eq. (21) are nothing more than hadronic currents. It is surprising how Fermi's theory showed such deep analogies with the modern theory of beta decay, despite the different physical-mathematical approach and the opposite starting hypotheses (such as the contact interaction rather than that mediated by a massive boson).

Let us see now how the operators $c_{ij}$ and $c_{ij}^{\dagger}$ act on the nucleon functions. To this purpose we denote the initial unperturbed state of the nucleus as:

$$(\rho, n, N_1, N_2, \cdots, N_i, \cdots, M_1, M_2, \cdots, M_j), \tag{22}$$

where $n$ is the quantum state of the nucleon. The occupation numbers $N_i$ and $M_j$ may assume values 0 and 1, respectively, and refer to the states occupied by the electron and neutrino. Denoting by $u_n(x)$ the neutron eigenfunction and by $v_n(x)$ that of the proton, applying the perturbation theory, we get the explicit form of the perturbation matrix:

$$H^{1mn\cdots 0_i \cdots M_1 \cdots 0_j}_{-1mN_1\cdots 1_i \cdots M_1\cdots 1_j} = \pm \int v_m^* c_{ij}^* u_n d\tau. \tag{23}$$

The integral is extended to the space of configurations of the heavy particle (except the internal coordinate $\rho$). Equation (23) represents the perturbation causing the *neutron* → *proton* transformation.

In the spectroscopic theory, where Fermi used *classical* quantum formalism, the perturbation Hamiltonian is:

$$H^{(s)}_{int.} = -e\frac{\hat{p}\hat{A} + \hat{A}\hat{p}}{mc}, \tag{24}$$

where $\hat{p}$ is the impulse operator acting on the coordinates of the electronic wavefunction and $\hat{A}$ the vector potential given by:

$$\hat{A}(x,t) = \sum_{k,\lambda}\left(\frac{\hbar}{2\varepsilon_0 \omega_k \Omega}\right)^{1/2}\left[a_{k\lambda}\varepsilon_{k\lambda}e^{i(k\cdot x - \omega_k t)} + a_{k\lambda}^{\dagger}\varepsilon_{k\lambda}^* e^{-i(k\cdot x - \omega_k t)}\right], \tag{25}$$

where $\omega_k$ is the frequency of the harmonic oscillator related to the photon and $\varepsilon_{k\lambda}$ its polarization vector. Since Hamiltonian Eq. (24) describes the phenomenon occurring in the atomic electron cloud, its values are significantly different from zero only in a region having the atomic range. Making explicit the impulse operator, Hamiltonian Eq. (24) may be rewritten as:

$$H^{(s)}_{int.} = e\frac{i\hbar}{2mc}(\nabla \cdot \hat{A} + \hat{A}\cdot \nabla). \tag{26}$$

Comparing Hamiltonian Eqs. (16) and (26), we see that the elementary charge $e$ represents the constant of the electromagnetic interaction, the $\nabla$ differential operator is analogous to the $Q$ operator and the vector potential $\hat{A}$ is analogous to that of the weak interaction. The electromagnetic perturbation matrix is:

$$H_{n,n_1,\cdots,m,m_1,\cdots} = \int \varphi_n^* u_1 u_2 \cdots H^{(s)}_{int.} \varphi_m u_{m_1} u_{m_2} \cdots dx dx_1 \cdots, \tag{27}$$

where $x$ is the vector of the electron coordinates, $x_n$ are the photon coordinates, $\varphi_n$ and $\varphi_m$ are the electron eigenfunctions of the initial and final states, and. $u_k$ and $u_{m_k}$ are the photon eigenfunctions corresponding to different energy states. Substituting Hamiltonian Eq. (26) in the integral Eq. (27), we get the explicit form of the matrix components:

$$H_{n,n_1,\cdots,m,m_1,\cdots} = -\frac{e}{m}\left(\frac{\hbar}{2\varepsilon_0\omega_k\Omega}\right)^{1/2}(\varepsilon_{k\lambda}\Box P_{knm})\left[\begin{array}{c}(n_k+1)^{1/2}\\ n_k^{1/2}\end{array}\right]_{m_k=n_k-1}^{m_k=n_k+1}. \quad (28)$$

The terms in the square brackets are the possible non zero values corresponding to the conditions $m_k = n_k \pm 1$. The term $P_{knm}$ is:

$$P_{knm} = -i\hbar \int \varphi_n^* sin(\boldsymbol{k}\Box\boldsymbol{x} - \omega_k t)\nabla\varphi_m d\boldsymbol{x}. \quad (29)$$

Therefore, unless numerical terms, the parallelism between the two theories may be summarized as follows:

$$\pm \int v_m^* c_{ij}^* u_n d\tau \equiv \int \varphi_n^* sin(\boldsymbol{k}\Box\boldsymbol{x} - \omega_k t)\nabla\varphi_m d\boldsymbol{x} . \quad (30)$$

The neutron and proton eigenfunctions are *equivalent* to different electronic states of an atom. This suggests that, if the spectroscopic process occurs by the annihilation (absorption) or creation (emission) of a photon, beta decay should follow a similar mechanism with the creation of a boson ($W^-$) which, unlike the photon, in turn, decays into two different particles (electron and antineutrino). It is possible that Fermi had evaluated this hypothesis without taking it into consideration, as it would have been a further speculative hypothesis to add to the others which were already very challenging for the physics of the period. The choice, therefore, was to proceed in formulating the theory of beta decay under a more *moderated* hypothesis of contact interaction.

## 4      Probability Amplitudes

Let us consider negative beta decay. Since the emitted light particles are free, the occupation numbers $N_s$ and $M_\sigma$ are both zero. At the time $t_0 = 0$, the system is formed by the neutron whose state $(1,n,0_s,0_\sigma)$ has unitary probability amplitude:

$$a_{1,n,0_s,0_\sigma} = 1. \quad (31)$$

The eigenfunction describing the neutron is $u_n(\boldsymbol{x})$. Considering a time small enough to hold that the condition in Eq. (31) is still valid, we can apply the time-dependent perturbation theory, getting:

$$\dot{a}_{-1,m,1_s,1_\sigma} = -\frac{2\pi i}{h}H_{-1,m,1_s,1_\sigma}^{1,n,0_s,0_\sigma} e^{\frac{2\pi i}{h}(-W+H_s+K_\sigma)t}, \quad (32)$$

where $H_s$ and $K_\sigma$ are, respectively, the energies of electron and neutrino. The hypothesis to consider a small enough time implies that the uncertainty on the energy is very large:

$$\delta E \geq \hbar/\delta t . \quad (33)$$

This is entirely consistent with the decay mechanism provided by the Standard Model, where the $W^-$ boson is created from the transformation of the quark down into up. The creation of the $W^-$ boson requires a lot of energy which is borrowed from the quantum vacuum [12]. Fermi did not know the mechanism of mediator bosons, but, implicitly and indirectly, his theory physically foresaw them. With integration Eq. (32), we get:

$$a_{-1,m,1_s,1_\sigma} = -H^{1,n,0_s,0_\sigma}_{-1,m,1_s,1_\sigma} \frac{e^{\frac{2\pi i}{h}(-W+H_s+K_\sigma)t}}{(-W+H_s+K_\sigma)}, \tag{34}$$

whose square modulus provides the probability of neutron decay.

Let us consider the spectroscopic. The perturbation theory leads to [9]:

$$\dot{a}_{n,n_1,\cdots} = -\frac{2\pi i}{h} \sum_{m,m_1,\cdots} a_{m,m_1,\cdots} H^{n,n_1,\cdots}_{m,m_1,\cdots} e^{-\frac{2\pi i}{h}\left[(E_m++h\nu_1 m_1+\cdots)-(E_n++h\nu_1 n_1+\cdots)\right]t} \tag{35}$$

The structure of Eq. (35) is very similar to Eq. (32) with the following analogies:

$$\begin{cases} (E_m - E_n) \equiv W \\ h\sum_i \nu_i (m_i - n_i) \equiv (H_s + K_\sigma) \\ \sum_{m,m_1,\cdots} a_{m,m_1,\cdots} H^{n,n_1,\cdots}_{m,m_1,\cdots} \equiv H^{1,n,0_s,0_\sigma}_{-1,m,1_s,1_\sigma} \end{cases}$$

The $W$ neutron-proton energy difference is analogous to the energy difference between the initial and final states of the atomic electron, while the energy difference among all the possible photon quantum states are analogous to the total energy of the beta decay light particles. The deep similarity between the two theories comes from the fact that both have been formulated using the perturbation theory, and this suggests that the two phenomena, apparently so different, are instead the result of the same mechanism. Perhaps, Fermi had guessed a way to unify the electromagnetic interaction with the weak one, using in a pragmatic, but ingenious way, the few germinal ideas on the atomic nucleus and on the neutrino existence.

## 5  Mean Lifetimes

Another point of similarity between the two theories is represented by the calculation of the mean lifetimes given by [5,9]:

$$\begin{cases} 1/\tau^{(s)} = e^2 \frac{64\pi^2 \nu^3}{3hc^3} |X_{mn}|^2 \\ \frac{1}{\tau^{(\beta)}} = g^2 \frac{(mc^2)^5}{2\pi^3 \hbar(\hbar c)^6} \left|H^{1,n,0_s,0_\sigma}_{-1,m,1_s,1_\sigma}\right|^2 \end{cases} \tag{36}$$

The structure of these two lifetimes is the same: the square of the coupling constants, a numerical term and the square of the perturbation matrix terms ($X_{mn}$ is the matrix representing the atomic electrical dipole). Considering that the emitted photon wavelength is:

$$\lambda = \frac{c}{\nu}, \tag{37}$$

and that the De Broglie wavelength of the electron is:

$$\lambda_e = \frac{h}{mc}, \tag{38}$$

the lifetimes Eq. (36) may be rewritten as:

$$\begin{cases} 1/\tau^{(s)} = e^2 \dfrac{64\pi^4}{3hc^3\lambda^3} |X_{mn}|^2 \\ 1/\tau^{(\beta)} = g^2 \dfrac{(mc^2)^2}{\hbar^4 c^3 \lambda_e^3} \left| H_{-1,m,1_s,1_\sigma}^{1,n,0_s,0_\sigma} \right|^2 \end{cases}. \tag{39}$$

Equation (39) has exactly the same algebraic structure, proving that the two processes occur with the same mechanism.

## 6  Generalized Fermi's Golden Rule

Fermi's golden rule leads to computing the probability that, in a given time, a transition from an initial to a final state occurs. It can be used both for spectroscopic [9] and beta decay [22] processes. The probability is:

$$P(t) = \frac{2\pi}{\hbar} \left| \langle \psi_f | H_{int.} | \psi_i \rangle \right|^2 \rho, \tag{40}$$

where $\rho$ is the (energy) density of the final state. In beta decay, where the interaction is at contact, the Hamiltonian can be written as:

$$H_{int.}^{(\beta)} = g\delta(r_e - r)\delta(r_\nu - r). \tag{41}$$

The integral of Eq. (40) becomes:

$$\langle \psi_f | H_{int.}^{(\beta)} | \psi_i \rangle = g \int \psi_e^*(r) \varphi_\nu^*(r) v_m^*(r) u_n(r) dr. \tag{42}$$

For the spectroscopic case, the Hamiltonian may be written as:

$$H_{int.}^{(s)} = e\delta(\nu_f - \nu), \tag{43}$$

where $\nu$ is the frequency of the photon.
The integral in Eq. (40) becomes:

$$\langle \psi_f | H_{int.}^{(s)} | \psi_i \rangle = g \int u_1 u_2 \cdots \varphi_n^*(r) u_{m_1} u_{m_2} \cdots \varphi_m(r) dr dx_1 dx_2 \cdots \tag{44}$$

The integral Eq. (44) is analogous to Eq. (42), also proving that, in Fermi's golden rule, the two theories are linked by the same physico-mathematical mechanism.

## 7  Discussion

The electroweak unification is the great success of the Standard Model, even if some non-secondary aspects of the theory remain to be clarified, such as the neutrino mass [23]. The electroweak theory is based on the symmetry principle, which refers to the work of Yang-Mills [24], consisting of the invariance of the theories under gauge transformations. In particular, the electroweak unification falls within the $SU(2) \times SU(1)$ group with symmetry breaking leading to the massive bosons $W^\pm$ and $Z$. But, in the years when Fermi formulated his theories, the formalism of the group theory, with the pioneering works of Weyl and Majorana [25-26], was not yet widespread among physicists and remained confined to a small group of theorists. Moreover, although Fermi had a solid mathematical basis, he had an experimental physics background, and his approach to the new questions of physics was mainly based on the physical meaning of the equations, rather than on their mere formalism. This explains why Fermi, in formulating his theories, took as a common starting point the fact that the energies to which the two phenomena occur, spectroscopic and beta decay are small compared to the respective unperturbed systems. So, as the (gauge) invariance of the Lagrangian density under the action of the symmetry group elements leads to the interaction mechanism mediated by vector bosons, the formalism of the perturbation theory leads to equations whose structures suggest that radiative and radioactive phenomena are intimately connected. In the previous sections, we have given evidence of the specularity between the Fermi approach and that of the Standard Model and, with the suitable comparison terms, we can affirm that the two Fermi theories were the first historical contribution to electroweak unification. Even if Fermi had guessed this possibility, the times were still too premature to be discussed in his work [5], which already contained strong speculative hypotheses that caused its rejection by the editor of *Nature*. The disappointment suffered by Fermi for this *failure*, although today his two theories are considered real milestones for rigorousness and innovation, turned him away from theoretical research to applied research, which resulted in the successes we all know. Undoubtedly, this was a case of serendipity, but probably did not allow Fermi to make that potential step forward to anticipate over time the final electroweak unification that, instead, had its consecration with the Standard Model.

## 8   Conclusion

The great affinity between the spectroscopic and beta decay theories is due to the fact that Fermi formulated the latter based on the mechanism with which the first takes place. This analogy has emerged also in the Lagrangian field. In fact, for the radiative processes, it is given by:

$$\mathrm{L}^{(s)} = e J^{(em)}_\mu A^\mu = e \left( \bar{\varphi}_n \gamma_\mu \varphi_m \right) A^\mu, \qquad (45)$$

while, for beta decay, it is:

$$\mathrm{L}^{(\beta)} = g J^{(n \to p)}_\mu J^\mu_{(\nu-e)} = g \left( \bar{v}_m \gamma_\mu u_m \right) \left( \bar{\psi}_e \gamma_\mu \varphi_\nu \right). \qquad (46)$$

The comparison between the two Lagrangians recalls the relation in Eq. (20) and proves that the mechanism of the two processes is the same. Fermi arrived at the same conclusion years after the publication of his theory on beta decay [27], just using the Lagrangian formalism of quantum fields. We can conclude that the electroweak unification finds its own origins precisely in the pioneering works of Fermi on beta decay and on its tight link with the spectroscopic theory. It is only by following this historical approach that the full physical meaning of unification emerges, which, on the contrary, tends to remain latent if it is dealt with using only the mathematical formalism that characterizes the Standard Model.